\documentclass[sigconf]{acmart}
\AtBeginDocument{%
  }

\setcopyright{none}
\copyrightyear{2026}
\acmYear{2026}
\acmDOI{XXXXXXX.XXXXXXX}
\acmConference[PREPRINT]{ICSE'26 Companion. Please note that this document is a preliminary version}{2026}{Accepted at ICSE'26 Track ``Software architecture BoF''}
\acmISBN{XXX-X-XXXX-XXXX-X/XXXX/XX}

\usepackage{todonotes}

\theoremstyle{definition}

\begin{document}

%\title{Towards "Creative Systems": the coming age of AI-enabled systems for quality achievement at runtime}
\title{Toward architecting self-coding information systems}


% \author{Rodrigo Falc\~{a}o}
% %\authornote{Both authors contributed equally to this research.}
% \email{rodrigo.falcao@iese.fraunhofer.de}
% \orcid{0000-0003-1222-0046}
% \author{Frank Elberzhager}
% %\authornotemark[1]
% \email{frank.elberzhager@iese.fraunhofer.de}
% \affiliation{%
%   \institution{Fraunhofer IESE}
%   \city{Kaiserslautern}
%  % \state{Ohio}
%   \country{Germany}
% }

\author{Rodrigo Falc\~{a}o}
\orcid{0000-0003-1222-0046}
\affiliation{%
  \institution{Fraunhofer IESE}
  \city{Kaiserslautern}
  \country{Germany}}
\email{rodrigo.falcao@iese.fraunhofer.de}

\author{Frank Elberzhager}
\orcid{0000-0002-8748-3927}
\affiliation{%
  \institution{Fraunhofer IESE}
  \city{Kaiserslautern}
  \country{Germany}
}
\email{frank.elberzhager@iese.fraunhofer.de}

\author{Karthik Vaidhyanathan}
\affiliation{%
  \institution{IIIT Hyderabad}
  \city{Hyderabad}
  \country{India}}
\email{karthik.vaidhyanathan@iiit.ac.in}

\renewcommand{\shortauthors}{Falc\~{a}o et al.}

\begin{abstract}

In this extended abstract, we propose a novel research topic in the field of agentic AI, which we refer to as self-coding information systems. These systems will be able to dynamically adapt their structure or behavior by evaluating potential adaptation decisions, generate source code, test, and (re)deploy their source code autonomously, at runtime, reducing the time to market of new features. Here we motivate the topic, provide a formal definition of self-coding information systems, discuss some expected impacts of the new technology, and indicate potential research directions.

\end{abstract}




\received{9 January 2026}
\received[revised]{XX January 2026}
\received[accepted]{XX January 2026}

\maketitle

\section{Motivation}\label{sec:motivation}

%Generative AI, in particular large language models (LLMs), has become an absolutely pervasive topic in software engineering. When it comes to software architecture, 

We have experienced unprecedented interest in artificial intelligence (AI) in the current decade. This can be asserted in the software engineering community, in general, and the software architecture community, specifically. Last year's ICSE has featured, once again, AI in multiple tracks, sessions, workshops, and keynotes\footnote{\url{https://conf.researchr.org/program/icse-2025/program-icse-2025/}}. The same can be observed in the software architecture community: ICSA 2025, held in Odense, had the theme ``Architecting for the next generation of intelligent systems''. The main conference featured AI in two keynotes and multiple sessions.

Generative AI, especially language models, plays an important role in the current ``AI spring''. These models have shown impressive capabilities in generating human-like content for virtually any task in any domain. Furthermore, base models can be (and have been) fine-tuned to excel on specific tasks such as writing code in a given programming language, automating configuration, text translation, text summarization, engaging in conversation, performing sentiment analysis, and doing image classification and generation, to name a few. This has allowed exploration of various use cases across domains. %Among the most popular LLM use cases is implementing specialized AI assistants (chatbots).

When applied to software architecture, generative AI can be viewed as a two-fold field. One is concerned about how generative AI can support software architects to perform architecture activities at development time (e.g., quality assessment of UML diagrams \cite{de2025llm}, generation of architectural design decisions \cite{dhar2024can}, documentation evaluation, code assistance, code generation, and test case creation  \cite{jahic2024state}).
On the other hand, there has been a growing interest in understanding how to architect systems that implement generative-AI-based components to achieve certain goals at runtime -- a comparatively less explored aspect of generative AI in software architectures~\cite{esposito2025generative}. We can divide this runtime aspect into two parts: using genAI to improve \textit{functional aspects} of the systems and using genAI components to improve \textit{quality aspects} of the systems. The former, on the one hand, has gained more attention, especially through the introduction of the ``retrieval augmented generation'' (RAG) in 2020 \cite{lewis2020retrieval}. From a purely conceptual perspective, RAG realized and improved a long-ago idea for ``intelligent information systems'' \cite{lebowitz1983intelligent}. Powered by LLMs, the concept thrived, and research on RAG has been frequently found, from reference architectures (e.g., \cite{xu2025design}) to practical use cases (e.g., \cite{adnan2025leveraging}).

On the other hand, research on genAI at runtime to address quality aspects is now taking its first steps. For example, a recent experiment with LLMs showed that they can be very effective in addressing interoperability \cite{falcao2025evaluating}. One of the strategies used was code generation \textit{at runtime} (which makes sense as LLM-based code has already been vastly adopted at design time to support architects and developers): in the experiment, the system used LLMs to generate code and, afterward, the code was cleaned, compiled, deployed, tested, and used to convert data from an unknown representation to a desired target representation -- all this on the fly. This experience contributes to a vision of what we regard as a coming research topic: the architecting of \textbf{self-coding information systems}.

% \inlinetodo{
% •	Status of development of [Gen]AI technology
% •	Its applications to and effectiveness in code-related tasks
% •	The increasing need for software-based autonomy
% •	An envisioned future: “creative software systems”* --> self-coding systems
% }

\section{Self-coding information systems}\label{sec:whatis}

% \begin{definition}
%     A self-coding information system is a software-based system that has the ability to implement functional requirements and/or address quality requirements by generating the source code of at least one of its subsystems and using the generated source code at runtime.
% \end{definition}
We define a self-coding information system as \textit{an information system whose software part can implement functional requirements or address quality requirements by generating the source code of at least one of its subsystems and using the generated source code at runtime}. At the core of the definition is the capability of these systems to generate (or regenerate) their source code. This idea builds on top of existing definitions that hint at autonomy, such as autonomous systems \cite{rasmus2023autonomous}, context-aware systems \cite{schilit1994context}, autonomic computing \cite{kephart2003vision} (in this case, the authors elaborate on a broad vision of self-* systems that includes autonomous modification of software, but in terms of applying software patches, not self-coding) and self-adaptive systems (see e.g., \cite{cheng2008software}, where the authors do mention adaptation that changes the source code but through dynamic source-code weaving, not actual code generation). While our definition does not prescribe how these systems achieve their self-coding capabilities, agents based on language models (especially those tuned for coding tasks) are nowadays the most straightforward option; therefore, we assume it in this paper.

Self-coding information systems show a high degree of modifiability. As defined by Bass et al. in \cite{bass2013software}, ``[m]odifiability is about change'' and must be planned by architects considering what can change, the likelihood of the change, when and by whom the change is made, and the change costs. In self-coding information systems, modifications can be done to the source code, at runtime (reduced time to market), by the system itself (autonomy), and cheaper as if it were done by humans. Conversely, in traditional software-based systems, configuration is a common practice for modifiability at runtime; however, it has a limited reach, for the more configurable a system is, the more complex its architecture becomes. Modifications that require change in the source code are traditionally performed at development time.

\section{Expected impact}

We focus on three aspects to classify the impact of the emergence and establishment of self-coding information systems: %into the following categories:

\paragraph{Architectural trade-offs} Certain architectural trade-offs for the improved modifiability might appear. For example, the analyzability of self-coding information systems, as self-generated code may be harder to understand; functional correctness can also come into question as language models are non-deterministic; resource utilization is increased by the energy consumption required for performing inferences; installability is also impacted as, depending on the size of the models and the system usage requirements, special hardware (e.g., GPUs) may be needed.

\paragraph{Software engineering roles} Existing roles must be adapted to this new class of systems. For example, requirements engineering and software architecture activities will expand to a metalevel: beyond specifying and implementing a certain desired function, software engineers will be concerned about specifying and implementing an autonomous component that can implement, test, and deploy a class of desired functions. Considering interoperability, for instance again, the LLM-based agent is not general purpose but tailored for the concrete and specific interoperability task. In some cases, user interfaces may include natural language interaction to trigger the self-coding capabilities (see, for instance, OpenAI's App SDK \cite{openai2025introducingApps}, which promises to allow users to ``talk'' with their apps).

\paragraph{Social and organizational aspects} Finally, social and organizational impacts may emerge from scenarios where self-coding information systems exist. On the one hand, companies might be able to evolve their systems faster. We assume that the level of autonomy will grow and consider more and more software system qualities, such as interoperability, maintainability, or testability. As the general idea is not restricted to a domain or sector, to a certain size of a software system, to a certain programming language, or something else, the general potential is rather high, and thus the impact. Productivity might grow significantly. On the other hand, would human developers become a luxury in small and medium-size organizations if self-coding information systems advanced? Would the role of software developer be (at least partially) replaced with an advanced user who could chat with the system to evolve its functionalities?

\section{Research directions for self-coding information systems}

% \inlinetodo{
% •	Challenges
%     - reliability
% •	Limitations
%     - doesnt work all the time
% •	Research directions
%     - reliability strategies
%     - architectural patterns
%     - self-architecting systems --> systems that are able to "explain themselves"?
%     - economic analysis of self-coding systems
% }

A big challenge in today's solution is the ``missing last step'', i.e., solutions with genAI often work, but the final reliability is often missing. This is usually not acceptable from a quality perspective, as it is unclear when the results are correct; respectively, this has to be found out with additional effort or mechanisms to compensate. Researching \textbf{reliability strategies} for self-coding information systems is an essential research direction.

%\paragraph{Architectural patterns}
\textbf{Reference architectures} must be designed (or abstracted from concrete architectures as patterns) describing common ways of realizing self-coding information systems. These patterns will be informative from different perspectives. For example, they can guide deployment decisions or help limit the scope of the self-coding functionalities.

%\paragraph{Maintainability}
Another aspect is how good self-coding information systems are considered from the perspective of \textbf{maintainability} in general. We face the situation that auto-generated code is sometimes difficult to understand, compromising the ability of a human developer to evolve or fix a bug in it. Furthermore, self-coding information systems are not exclusively self-coded, i.e., human developers will also change the code. Therefore, proper workflows to support the coexistence with human developers are needed. Note that this goes beyond current practices involving the usage of coding assistants, as they are not part of the target system and do not update the target system at runtime.

As self-coding information systems require special hardware, and also considering the inference costs, it is necessary to develop simple ways of determining whether a self-coding capability is adequate or is inadequate for a certain aspect of a given system from an \textbf{economic perspective}. Returning to the interoperability example of Section~\ref{sec:motivation}: given that a self-coding component and a team of developers would both produce the same functional correctness outcome, which strategy would be more economical? Several variables can play a role in such an equation.

%\paragraph{Self-architecting}
Finally, self-coding information systems could evolve to become not only self-coding but also \textbf{self-architecting systems}, which, for example, document their design decisions and at least partially could also review their architectures.

%Furthermore, it is still unclear what is the ideal context that we should provide genAI in order to fulfill the tasks at its best. 

\begin{acks}
To our colleagues Rafael Capilla, Pablo Antonino, and Ivan Compagnucci, whose ideas in the co-organization of the SAGAI workshop have contributed to this manuscript (R.F. and F.E.). AI tools were used for proofreading.
\end{acks}

\bibliographystyle{ACM-Reference-Format}
\bibliography{refs}


\end{document}